# Degradation and Failure Mechanisms of Complex Systems: Principles

Tarannom Parhizkar[1a], Theresa Stewart[a], Lixian Huang[a], Ali Mosleh[a]

B. John Garrick Institute for the Risk Sciences, University of California, Los Angeles, 404 Westwood Plaza, Los Angeles, CA 90095, USA

**Abstract**

A cyber physical human complex system failure prevents the accomplishment of the system's intended function. The failure of a complex system could be a breakdown of any system hardware, human related factors, application software, or the interaction between these components. Having knowledge about all these three components would allow us to better understand the behavior, interactions, and the associated failure mechanisms of the cyber physical human systems as a whole.

In this study, degradation mechanisms in these three components are classified and discussed. The main categories are hardware related degradation mechanisms including mechanical, thermal, chemical, electronic and radiation effects degradation mechanisms. In addition to hardware related degradation mechanisms, human failure modes, software errors, and the failures due to cyber-physical-human interactions are presented and discussed. This paper covers the main types of failure mechanisms in complex systems and is beneficial for developing conceptual risk and reliability models for complex systems.

---

[1] Corresponding author

1 Introduction

Complex system components degrade over time due to different degradation mechanisms, [1]. Degraded components can negatively affect the system performance, reduce the system lifetime and even result in a catastrophic failure, [2]. In addition, degradation analysis can be used to assess reliability when few or even no failures are expected in a life test.

Degradation mechanisms in a complex system depend on many different factors such as type of the system, components material, system application, operating and environmental conditions, [3], [4], [5]. In complex systems, we have cyber-physical-human interactions, [6], [7]. A cyber physical human complex system is a system that is made of interacting components of software, hardware, and human operators. The three different elements in cyber physical human systems are hardware, software and humans that have different failure behavior characteristics. It is important to understand the difference between them and their failure modes, [8]. However, more importantly, having knowledge about all these three components would also allow us to better understand the behavior, interactions, and the associated failure mechanisms of the cyber physical human systems as a whole.

In this paper, hardware degradation including mechanical, thermal, chemical degradation mechanisms, degradation of electronic devices, radiation effects degradation mechanisms are introduced and discussed.

In this study, degradation mechanisms are categorized into eight main types. The main categories are hardware related degradation mechanisms including mechanical, thermal, chemical, electronic and radiation effects degradation mechanisms, discussed in Section 2 to Section 5, respectively. The mechanical, thermal, chemical degradation mechanisms are classified to wear-out and overstress failures. Wear out mechanisms are degradation mechanisms that happen gradually in the component and result in system aging and performance deterioration. As time passes, the degradation mechanism will cause exceedance of system threshold and system failure. Overstress mechanisms, on the other hand, are sudden degradation mechanisms. In this type of mechanism, operating and environmental conditions are out of nominal range of system operating conditions and results in system failure. This paper will discuss both wear out and overstress failure mechanisms of mechanical, thermal, and chemical natures. Due to the complexity in electronics, the degradation and failure mechanisms are introduced based on different physics causes.

In addition to hardware related degradation mechanisms, we could have software and human errors resulting in system failure that are discussed in Section 4 and 5, respectively. Finally, failure mechanisms due to components interaction are introduced. The interactions result in exchange of matter, energy, force, and/or information, and we can have all the combinations between hardware-software, hardware-human and the human-software, but also the intersection between all three. In Section 8, different types of interaction failure mechanisms are presented and discussed. It should be noted that this work is not intended to give the reader a thorough understanding of each individual failure mechanism, but rather is intended to introduce the reader to the types of failure or degradation that may be encountered in a wide variety of applications and environments.

## 2 Mechanical degradation and failure mechanisms

### 2.1 Wear out mechanisms

Mechanical wear could be defined as a process of progressive removal of material from a solid surface while it is in moving contact with another solid, liquid, or gaseous substance [9].

#### 2.1.1 Abrasion

Abrasion wear is one of the most common types of wear and appears across many applications accountings for more than 50% of all wear-related failures in industrial equipment, costing between 1 and 4% of the gross national product of industrialized nation [10] [11].

#### 2.1.2 Adhesion

Adhesive wear is another common type of wear that occurs in systems. It occurs whenever two solid surfaces have rubbing contact and even if all wear mitigation plans have been implemented, this type of wear will remain [12].

#### 2.1.3 Surface fatigue

This type of wear occurs when two surfaces have rolling contact. Wear here is different from sliding surfaces. As the rolling body rotates, the shear stress ranges from zero to a maximum value, resulting in cyclic stress in the components [13].

### 2.1.4 Erosive wear

In this type of wear, some particles slide and roll against the surface. The particles could have three different speed levels. At low speed, we have abrasion wear, and if these particles touch the surface on a cyclic basis, then it will result in fatigue wear out. At medium speed, plastic deformation or erosion by brittle fracture will occur. At higher speed, melting will occur [14].

### 2.1.5 Fretting wear

Fretting is a type of wear that occurs in systems with two parts in close contact, vibrating at small amplitudes [15]. This type of wear out mechanism occurs commonly in components that contain parts with relative motion such as bearings.

### 2.1.6 Creep deformation

Creep deformation describes the process by which solid materials undergo gradual plastic (permanent) deformation at stresses below the yield stress. Creep is a time dependent process and the rate of deformation from creep increases with temperature and stress [16] [17].

### 2.1.7 Fatigue

Fatigue failure occurs as a result of long-term cyclic loading within the design limits of a component. Under these loads, initial defects may become crack initiation sites which grow as the region is loaded and unloaded [18] [19].

### 2.1.8 Cavitation pitting

Cavitation pitting is a type of pitting fatigue as a result of vibration and movement of liquids in contact with solids. It occurs mostly at low pressure regions where voids can form. Any moving system in liquid will experience cavitation pitting [20].

### 2.1.9 Resonance disaster

When a structure encounters oscillations from wind, earthquakes, or vibrating motions that match the structure's natural frequency of vibration, it is able to pick up motion very easily and may sway violently or fracture. Resonance as a phenomenon is one that is often used in clocks to maintain motion of the timekeeping mechanism, such as a pendulum [21].

## 2.2 Overstress mechanisms

Overstress failure occurs when a material is subjected to unusually harsh conditions that aren't expected in the design of the part [22] [23] [24].

### 2.2.1 Plastic deformation

Whenever a solid material is subjected to a load, it will deform as a result of that load. When the load is relatively small, this deformation, referred to as elastic deformation, is reversible [25].

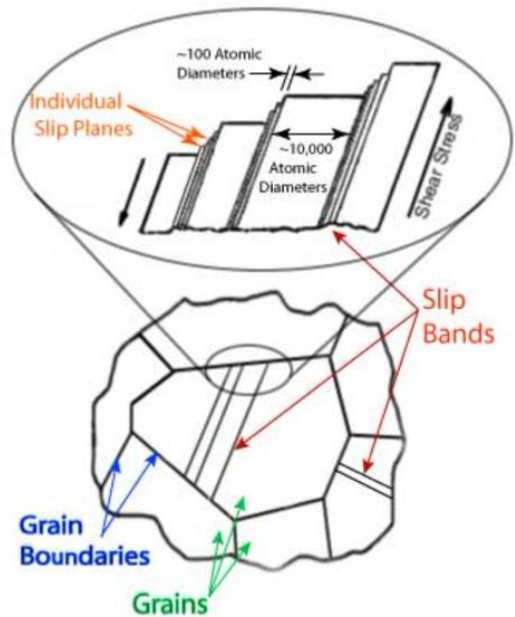

*Figure 1. Slip bands in plastic deformation process*

### 2.2.2  Fracture

Fracture is defined by the formation of new surfaces on a formerly intact piece of material not by growth but by damage from stress. Most often this occurs by the formation of cracks throughout the surface or body of the part, but in extreme cases can result in a part breaking into two or more pieces [16] [26].

### 2.2.3  Delamination

Composite materials, whether manmade or natural (such as woods), have anisotropic properties and they are more likely to fail in one loading direction over another [27].

### 2.2.4 Degradation due to residual mechanical stress

When a material undergoes plastic deformation (temperature gradients from thermal cycling, or structural changes from a phase transformation), stress may remain due to uneven changes in the dimensions of the part. These stresses are called residual stresses [25].

## 3 Thermal degradation and failure mechanisms

### 3.1 Wear out mechanisms

#### 3.1.1 Intermetallic Growth

Soldering in electronic components provides a combination of electrical connection and mechanical support, and thus the properties of these solder joints is very important to the performance of the component [28].

#### 3.1.2 Hillocks

An important source of mechanical stress in thin films is the thermal mismatch between the film and the substrate material. Depending on the sign of the mismatch and of the temperature change, tensile or compressive stresses can develop in the film. One mechanism of compressive stress relaxation which is specific to thin films is the formation of hillocks [29] [30] [31].

### 3.2 Overstress mechanisms

#### 3.2.1 Ductile to brittle transition

A ductile-to-brittle transition (DBT) is the point or range at which a material that is relatively ductile at high temperatures becomes more brittle at low temperatures [32] [33] [34] [35].

### 3.2.2 Glass Transition Temperature

Glass transition is a second-order transformation which occurs in amorphous materials - namely, polymers, [32]. This transition occurs over a range of temperatures and marks the point at which the polymer chain is able to move with increasing degrees of freedom without breaking.

### 3.2.3 Thermal shock

Thermal shock is the cracking mechanism as a result of rapid temperature change. It affects material properties such as toughness, thermal conductivity, and thermal expansion. In addition, as different parts expand by different amounts – creating a thermal gradient - it results in a differential strain which creates stress in the material. If this stress grows large enough, it can cause cracks to form [36].

### 3.2.4 Melting

Melting is the phase change of a substance from solid to liquid. It is a first-order transformation that happens at a constant temperature and results in increased molecular vibration [28].

### 3.2.5 Fatigue

Similar to mechanical fatigue, thermal fatigue is the weakening of a material caused by cyclic thermal loading resulting in crack growth and structural damage. Thermal fatigue is differentiated from mechanical fatigue in that it may occur by only thermal cycling, without mechanical loads. A key example of this is in turbines, where temperature differentials from the starting and stopping of the turbine produce temperature gradients in the material that leads to thermal fatigue, [16].

### 3.2.6 Degradation due to residual thermal stress

Residual thermal stresses are stresses introduced when a material undergoes heating or cooling and attempts to expand or contract. This can occur in one of two ways, [16] [37].

## 4 Chemical degradation and failure mechanisms

### 4.1 Wear out mechanisms

In this section, different types of chemical wear out mechanisms are explained. One of the main materials that suffer from chemical degradation is polymer. Polymers are a wide category of materials which includes synthetic thermoplastic and thermoset polymers, naturally occurring polymers present in organic material, and natural and synthetic rubbers.

#### 4.1.1 Polymer degradation-depolymerization

Depolymerization is a process which occurs at high temperatures and is defined as the decomposition of a polymer into one or more monomers, [38] [39] [40] [41].

#### 4.1.2 Polymer degradation – thermal oxidative degradation

Polymers can degrade by chain scission in which a long main chain breaks up into smaller chains of a smaller molecular weight. This type of degradation falls under the general category of thermal-oxidative degradation and is usually caused by exposure to heat or UV radiation while in the presence of oxygen [42].

### 4.1.3 Polymer degradation – Hydrolytic degradation

Many polymers are not impermeable to liquids and may either swell or dissolve. The ability for the polymer to take up a liquid depends on the solubility of the polymer and liquid. For polymers, the solubility is typically much higher with organic solvents than water [42].

### 4.1.4 Polymer degradation – Ozone Degradation

Polymers – especially those with doubly bonded carbon atoms in the main chain – are susceptible to chain scission caused or accelerated by reactions with ozone, as well as oxygen, to a lesser extent [42].

### 4.1.5 Corrosion

Corrosion is the process of metal breakdown by reacting with substances in their environment. One of the well-known examples is rusting, where iron reacts with oxygen and water from the environment to form hydrated iron oxides which are more commonly known as rust [42].

### 4.1.6 Outgassing

Outgassing is the process by which a material loses mass in very low-pressure conditions, most commonly in the vacuum of space [43].

## 4.2 Overstress mechanisms

Overstress chemical degradation mechanisms are chemical degradation mechanisms that results in material break down.

# 5 Electronics degradation and failure mechanisms

Electronic components can be divided into three categories: active, passive, and electromechanical. Active components include transistors, diodes, displays, etc. which can be used to amplify signal and power, and whose functions depend on external energy input.

## 5.1 Dielectric breakdown

Dielectric breakdown can be divided into five stages as shown in **Error! Reference source not found.**. The schematics of how defect clustering affects the leakage current evolution are shown in **Error! Reference source not found.**. There is a maximum electric field that the dielectric material can tolerate [44] [45] [46] [47] [48] [49] [50].

## 5.2 Bias temperature instability (BTI)

Bias temperature instability (BTI) is a degradation mechanism of MOSFETs. A threshold voltage shift may happen due to BTI when the MOS gate is applied a bias voltage under elevated temperature [51] [52] [53] [54] [55].

## 5.3 Hot carrier injection (HCI)

Hot carrier injection is a wear-out mechanism in MOSFETs. Hot carriers refer to the carriers that are accelerated under high electric fields and build up enough kinetic energy to inject into 'forbidden' regions like the gate oxide [56] [57] [58] [59] [60] [61].

### 5.3.1 Channel hot electron (CHE) injection

It usually happens when some electrons gain sufficient energy to pass the Si/SiO2 barrier at the drain end of the channel of the transistor without energy loss after collisions with channel atoms, [62] [63].

### 5.3.2 Substrate hot electron (SHE) injection

This mechanism happens when the transistor bulk is applied with a very large positive or negative bias. The carriers in the substrate are driven to the Si/SiO2 interface, gaining sufficient kinetic energy to by-pass the interface barrier, and injecting to the oxide [62].

### 5.3.3 Drain avalanche hot carrier injection (DAHC)

Under high drain voltage and low gate voltage, impact ionization of the channel current near the transistor drain can create electron-hole pairs [62]. It can also be called as avalanche multiplication which leads to drain avalanche hot carrier generation (DAHC).

### 5.3.4 Secondary generated hot electron injection (SGHE)

Impact ionization[2] generates secondary carriers. These carriers can be generated under DAHC or bremsstrahlung radiation near the drain, gaining sufficient kinetic energy and overcoming the surface energy barrier, that results in secondary generated hot electron injection [62].

---

[2] Happens when there is a sufficient large, applied voltage.

## 5.4 Electromagnetic interference (EMI)

Electromagnetic interference (EMI) is the disturbance caused by external sources to the target electrical circuits. External sources can cause disturbance on the target electrical circuits because of EMI. Such disturbance can be conductive, capacitive, magnetic, or radiative between the source and the victim circuits, [62].

## 5.5 Electrostatic discharge (ESD)

When two charged objects come close to each other, electrostatic discharge (ESD) can happen and generate electricity flow suddenly. Triboelectric charging, electrostatic induction, and energetic particles can cause ESD [64].

## 5.6 Electrical overstress (EOS)

Electrical overstress (EOS) can be induced by temperature, electric field, and electromigration. EOS can cause thermal runaway, which results in local thermal conductivity loss and producing more heat locally, [65] [66] [67].

## 5.7 Electromigration (EM)

Electromigration involves the metal ion migration due to momentum transfers under electrical stresses [62] [68].

## 5.8 Tin whiskers

Contact resistance in electronics is carefully controlled to ensure the device's functionalities. The contact degradation can be mechanical-induced, thermal-induced and electrical-induced. For example, similar to packaging failure, thermal expansion mismatch can induce internal mechanical stress and lead to fatigue cracking [69], but the sizes can vary depending on the cases, [70] [71] [72].

## 5.9 Self-healing accumulation

When overvoltage is applied on non-metallized film capacitors, small localized dielectric breakdown can occur and short the electrodes. However, in metallized film capacitors, since the electrode foils are thin, the foils can vaporize under high energy density in the fault area and avoid breakdown. This is the self-healing phenomenon, [73] [74].

## 5.10 Self-heating

The self-heating in the metallized film capacitors happens when body generated power exceeds the surface power dissipation capability. Self-heating raises the temperature of the capacitor, reduces the breakdown voltage, and even melts the capacitor, [75].

## 5.11 Electrochemical corrosion

For metallized film capacitors, the electrochemical corrosion of the thin electrode can be caused by the high ripple current and voltage, [76].

## 5.12 Thermal overstress induced electrolyte evaporation

Electrolyte evaporation is the primary wear-out mechanism in electrolyte capacitors. This mechanism is caused by high temperatures within the capacitor core, [76].

# 6 Radiation degradation and failure mechanisms

Radiation is defined as emission or transmission of energy through space or a medium. The word radiation came up from waves radiating that moves outward in every direction from a source, [77] [78].

## 6.1 Lattice displacement

Threshold displacement energy (TDE) is the most common quantity for describing radiation damage in materials. Thus, displacement energy is defined as the minimum required energy to create a stable defect in a material [79] [80] [81].

## 6.2 Ionization effects

Ionization effects result from the emission of energy from a medium through space or a medium that is sufficient to ionize molecules or atoms. The common sources of the ionizing radiation are X-rays, Alpha, Beta, Gamma, and Neutron radiations, [82] [83].

### 6.2.1 Non-destructive Single Event Effects

#### 6.2.1.1 *Single Event Upsets (SEU)*

Single event upsets occur in memories and sequential logic circuits. A single event upset causes changes of bits on a memory array, i.e., data loss. It can happen at any time, either at reading, writing, or the idle state of a circuit. A re-write can correct the error. This effect can show up as Multiple Bit Upsets caused by a single ion, [84].

### 6.2.1.2 Single Event Functional Interrupts (SEFI)

A single event upset that leads to temporary loss of device functionality is considered as a single event functional interrupt (SEFI), [85] [86].

### 6.2.1.3 Single Event Transients (SET)

In single event transients (SETs), an ion hit caused by cosmic rays leads to a voltage transient in the circuit. It is similar to electrostatic discharge and occurs in logic circuits, [87].

## 6.2.2 Destructive Single Event Effects

### 6.2.2.1 Single Event Latch-up (SEL)

A single event latch-up (SEL) occurs due to a low-impedance path generated by heavy ions or protons between two inner-transistor junctions, [88]. SEL is a strongly temperature-dependent event, [89].

### 6.2.2.2 Single Event Gate Rupture (SEGR)

Under a high electric field between gate silicon, heavy ions passing through the gate oxide will leave a path to discharge the capacitor. This process causes local overheating and microscopic explosion of the gate dielectric region and leads to permanent damage to gate dielectric, [90].

### 6.2.2.3 Single Event Burn-out (SEB)

A single event burn-out (SEB) occurs when the substrate under the source region gets forward-biased and the drain-source voltage is higher than the breakdown voltage of the parasitic structures.

This process requires a sufficient magnitude of current to overheat the device to destruction. It occurs often in bipolar and MOSFET power transistors and leads to permanent damage, [91].

#### 6.2.2.4 *Single Event Hard Errors (SEHE)*

A single event hard error (SEHE) occurs when heavy ions strike on memory arrays, leading to a large energy transfer. It causes an unchangeable state, called a stuck bit in memory, and leads to permanent damage, [92] [93].

## 7 Human failure modes

Human failure modes can be classified to human error, and violation. These two failures are covered in more detail in the following sections, [94].

### 7.1 Human error

Human error is defined as an unintentional decision or action and can be classified to slips and lapses (skill-based errors), and mistakes. Human errors are possible to happen to even the well-trained and/or experienced person, [94].

#### 7.1.1 Skill based errors (slip and lapse)

Slips and lapses occur in tasks that we can perform with low conscious attention; for instance, driving a vehicle. The tasks with low conscious attention requirements are very vulnerable to slips and lapses when the attention is diverted for a very short time [94].

### 7.1.2 Mistakes

Mistakes are defined as decision-making failures., i.e., It happens when a person performs a wrong task, believing it to be right. There are two main types of mistakes, including knowledge-based and rule-based mistakes, [94].

## 7.2 Violation

Violations are intentional failures and deliberately performing the wrong task. The violation of health and safety principles and rules is one of the biggest causes of injuries and accidents at workplaces, [94].

## 8 Software errors and failure mechanisms

Software errors and failure mechanisms can be classified to function failure mode, value-related and timing failure modes, and interaction failure modes, [95] [96] [97].

## 9 Cyber-Physical-Human (CPH) Systems' Interaction Failure Mechanisms

A cyber physical Human complex system is a system that is made of interacting components of software, hardware, and human operators (Figure 1). In these types of systems, we have cyber-physical-human interactions, [98], [99].

## 10 Discussion and conclusion

For the purpose of reliability analysis of complex systems, proper understanding of cross-filed degradation and failure mechanisms is essential. In many cases, various degradation mechanisms can have certain dependencies and consequently, the initiation of one degradation mechanism can positively or negatively affect the propagation of the others in the same system. An accurate reliability estimation of one complex system requires engineers not only to investigate the material, human and software reliability factors, but also to understand the interactions and dependencies among them.

In this study, different types of material degradation mechanisms in complex systems are classified and presented. In addition to material degradation, human errors, software failure, and cyber-physical-human interaction failure are presented and discussed.

The mechanism-based reliability studies are capable of facilitating the engineers to monitor the system health, help decision makers to weigh among reliability factors and even provide guidance on the selections of materials, system structures, operation conditions, operating strategies, operators and software.